\documentclass[twocolumn]{aastex63}
\usepackage{CJKutf8}
\usepackage{amsmath}

\received{}
\revised{}
\accepted{}

\shorttitle{OSSOS Neptune Trojans}
\shortauthors{Lin et al.}

\begin{document}

\title{OSSOS: The Eccentricity and Inclination Distributions of the Stable Neptunian Trojans}

\author[0000-0001-7737-6784]{Hsing~Wen~Lin (\begin{CJK*}{UTF8}{bkai} 林省文\end{CJK*})}
\affiliation{Department of Physics, University of Michigan, Ann Arbor, MI 48109, USA}
\correspondingauthor{Hsing~Wen~Lin}
\email{hsingwel@umich.edu}

\author[0000-0001-7244-6069]{Ying-Tung Chen (\begin{CJK*}{UTF8}{bkai} 陳英同\end{CJK*})}
\affiliation{Institute of Astronomy and Astrophysics, Academia Sinica, P. O. Box 23-141, Taipei 106, Taiwan}

\author[0000-0001-8736-236X]{Kathryn Volk}
\affiliation{Lunar and Planetary Laboratory, University of Arizona, 1629 E University Blvd, Tucson, AZ 85721, United States}

\author{Brett Gladman}
\affiliation{Department of Physics and Astronomy, University of British Columbia, 6224 Agricultural Road, Vancouver, BC V6T 1Z1, Canada}

\author{Ruth Murray-Clay}
\affiliation{Department of Physics, University of California Santa Barbara, USA}

\author[0000-0003-4143-8589]{Mike Alexandersen}
\affiliation{Minor Planet Center, Harvard-Smithsonian Center for Astrophysics, 60 Garden Street, Cambridge, MA 02138, USA}

\author[0000-0003-3257-4490]{Michele T. Bannister}
\affiliation{School of Physical and Chemical Sciences – Te Kura Mat\={u}, University of Canterbury, Private Bag 4800, Christchurch 8140, New Zealand}
\affiliation{Astrophysics Research Centre, Queen’s University Belfast, Belfast BT7 1NN, United Kingdom}

\author[0000-0001-5368-386X]{Samantha M. Lawler}
\affiliation{Campion College and the Department of Physics, University of Regina, Regina, SK S4S 0A2, Canada}

\author{Wing-Huen Ip (\begin{CJK*}{UTF8}{bkai} 葉永烜\end{CJK*})}
\affiliation{Institute of Astronomy, National Central University, 32001, Taiwan}
\affiliation{Space Science Institute, Macau University of Science and Technology, Macau}

\author[0000-0003-0926-2448]{Patryk Sofia Lykawka}
\affiliation{School of Interdisciplinary Social and Human Sciences, Kindai University, Japan}

\author[0000-0001-7032-5255]{J. J. Kavelaars}
\affiliation{NRC-Herzberg Astronomy and Astrophysics, National Research Council of Canada, 5071 West Saanich Rd, Victoria, British Columbia V9E 2E7, Canada}
\affiliation{Department of Physics and Astronomy, University of Victoria, Elliott Building, 3800 Finnerty Rd, Victoria, BC V8P 5C2, Canada}

\author{Stephen D. J. Gwyn}
\affiliation{Department of Physics and Astronomy, University of Victoria, Elliott Building, 3800 Finnerty Rd, Victoria, BC V8P 5C2, Canada}

\author{Jean-Marc Petit}
\affiliation{Institut UTINAM UMR6213, CNRS, Univ. Bourgogne Franche-Comté, OSU Theta F25000 Besançon, France}

\begin{abstract}

The minor planets on orbits that are dynamically stable in Neptune's 1:1 resonance on Gyr timescales were likely em:laced by Neptune's outward migration. 
We explore the intrinsic libration amplitude, eccentricity, and inclination distribution of Neptune's stable Trojans, using the detections and survey efficiency of the Outer Solar System Origins Survey (OSSOS) and Pan-STARRS1.
We find that the libration amplitude of the stable Neptunian Trojan population can be well modeled as a Rayleigh distribution with a libration amplitude width $\sigma_{A_{\phi}}$
of 15\degr.
When taken as a whole, the Neptune Trojan population can be acceptably modeled with a Rayleigh eccentricity distribution of width ${\sigma_e}$ of 0.045 and a typical sin(i) $\times$ Gaussian inclination distribution with a width ${\sigma_i}$ of $14\pm2^\circ$; however, these distributions are only marginally acceptable. This is likely because, even after accounting for survey detection biases, the known large (H$_r<8$) and small (H$_r \geq 8$) Neptune Trojans appear to have markedly different eccentricities and inclinations. 
We propose that like the classical Kuiper belt, the stable intrinsic Neptunian Trojan population have dynamically `hot' and dynamically `cold' components to its eccentricity/inclination distribution, with $\sigma_{e-cold} \sim 0.02$/$\sigma_{i-cold} \sim 6^{\circ}$ and $\sigma_{e-hot} \sim 0.05$/$\sigma_{i-hot} \sim 18^{\circ}$. 
In this scenario, the `cold' L4 Neptunian Trojan population lacks the $H_r \geq 8.0$ members and has $13^{+11}_{-6}$ `cold' Trojans with $H_r < 8.0$. On the other hand, the `hot' L4 Neptunian Trojan population has $136^{+84}_{-75}$ Trojans with $H_r < 10$ --- a population 2.4 times greater than that of the L4 Jovian Trojans in the same luminosity range.

\end{abstract}


\section{Introduction} \label{sec:intro}
The Neptunian Trojans (NTs) are minor planets that co-orbit with Neptune at semi-major axes $\sim30.1$ au. 
These objects librate in 1:1 resonance.
Like Neptune's other n:1 resonators, the 1:1 resonances also contain symmetric and asymmetric libration islands \citep{bea94, mal96, mur05, gla12, pik15, vol18, che19}. Asymmetric
1:1 librators are termed Trojans and librate around Neptune's L4 (leading) and L5 (trailing) Lagrange points; the extent of the stable region around each point depends on both orbital inclination and eccentricity. 
\citet{zho09, zho11} showed that NTs can be dynamically stable for billion years even at very high orbital inclinations ($> 25^{\circ}$). 

The symmetric librators have horseshoe co-orbital motion, encompassing both the L4 and L5 Lagrange points, but they are generally not long-term stable \citep[e.g.][]{bra04a, fue12}. 
Additionally, some known unstable NTs librate around their Lagrange point for timescales of only Myr or shorter, suggesting that they are temporarily captured and not ancient \citep{gua12, hor12b}. The existence of temporary NTs is expected \citep{hor12a}. Numerical simulations show that several percent of the Centaur population can be sticking to the 1:1 co-orbital resonances of Neptune and Uranus at any given time \citep{ale13}. 

In contrast, stable Trojans, i.e. asymmetric librators with Gyr-long dynamical lifetimes, are part of a population that dates back to events early in Solar System history \citep{lyk11, par15, gom16}. Many numerical studies suggest that during Neptune's outward migration, the initial NT population could be captured into its current orbits from a minor planet population that was previously excited to a wide range in eccentricities (e), inclinations (i), and libration amplitudes (A), which may explain the observed population \citep{nes09, lyk09, lyk10, par15, che16, gom16}. Furthermore, analysis of the long-term behaviour of these captured NTs revealed that they are not expected to
change their e, i, A significantly over Gyr timescales. In other words, once an object is captured as a NT, it can hold some memory of its primordial e/i at the time of capture. The same is true for NTs that may have formed in-situ \citep{lyk09, lyk11}. Thus, these
facts imply that the e-/i-distributions of currently known NTs can probe the conditions of the primordial population of NTs and place insightful constraints on cosmogonic models of the outer solar system.

Thirteen of the currently known 23 NTs are dynamically stable, maintaining Trojan behavior for more than 1 Gyr forward integrations \citep[see Table~\ref{tab:orbits}; c.f.][]{lyk11, par13, ale16, ger16, lin16, wu19}. 
Using stable NTs to identify the present-day NT e-/i-distributions offers constraints on these migration models and the possible origin of NTs. For example, \citet{lyk09, lyk11} predicted that stable NTs possess similarly wide e-distributions for captured objects and colder e-distributions for objects formed locally.

The sensitivity of Solar System surveys to the NT's e, i, and A distribution is a function of their sky coverage \citep{lin16, lin19}. For example, a low-inclination orbit will spend its orbital period entirely within a low-ecliptic-latitude sky, which is the predominant survey coverage for general minor planet surveys.
More inclined orbits spend a smaller fraction of their orbital period within low-ecliptic-latitude fields, and thus geometrically lower the likelihood of their population's detection.
Additionally, only Neptune's L4 point has been well targeted by surveys, due to the overlap of the L5 point on the Galactic plane in the era of digital sky surveys.
Merely three NTs are known from L5, while twenty are known from L4 (Table~\ref{tab:orbits}).

Past NT surveys have suggested that the current NT population has a broad inclination distribution. 
\citet{she06} surveyed near L4 and found a stable high-inclination NT; based on this discovery, they concluded that the population has a thick-disk distribution.
\citet{par15} noted that the eight then-known stable NTs mostly had high orbital inclinations, despite their detection in a variety of low-ecliptic-latitude surveys, such as \citet{she10, she10b} and \citet{par13}.
A statistical method has been applied in \citet{par15} to de-bias the observed distributions of orbital inclinations, eccentricities, and libration amplitudes. Typically, the intrinsic inclination distribution of outer minor planet populations is modeled using a sin(i) $\times$ Gaussian distribution \citep{bro01}. \citet{par15} confirmed the broad distribution, finding a width of $\sigma _i > 11^\circ$, with statistically inconclusive hints of bimodality. The broad inclination distribution would indicate a primordial NT cloud at least as thick as the Jovian Trojan cloud, the other major Trojan population in the Solar System. 

Motivation in the understanding of the NT inclination distribution became apparent following the NT observations of the Pan-STARRS1 survey \citep[PS1;][]{cha16} and the initial results of the Dark Energy Survey \citep{lin19}.
Both PS1 and the Dark Energy Survey have extensive sky coverage with large areas at high ecliptic latitudes, making them thoroughly sensitive to NTs on highly inclined orbits. 
In the PS1 analysis of \citet{lin16}, six stable L4 NTs were detected --- yet only one had an inclination larger than 20 degrees. 
As a raw observational result, this seems to suggest a colder NT inclination distribution than that found by \citet{par15}. 
\citet{lin16} found that the acceptable range of inclination widths, $ 7^\circ < \sigma _i < 27^\circ$, still generally agreed with \citet{par15}'s result of $\sigma _i > 11^\circ$.
On the other hand, the first-pass analysis of part of the Dark Energy Survey (DES) by \citet{lin19} detected five NTs in high-latitude fields, and two of the five NTs have $> 30^\circ$ inclinations. After removing the observation bias of DES using a DES survey simulator \citep{ham19}, \citet{lin19} found that this result indicates a high inclination widths ($\sigma _i \sim 26^\circ$) NT population. A full search of that survey for Neptunian Trojans is ongoing \citep{bern19}. 

Additional NT discoveries and non-detections from surveys with well-characterized detection biases can improve our understanding of the intrinsic NT inclination distribution. The Outer Solar System Origins Survey (OSSOS; \citet{ban16, ban18}) discovered four NTs. As OSSOS has highly quantified detection efficiencies, its survey biases in orbital parameter space can be thoroughly modeled, using a survey simulator \citep{law18a}. 
The smaller survey of \citet[A16;][]{ale16} discovered an additional stable NT, 2012 UV$_{177}$. 
A useful non-detection constraint comes from the sky covered by the related Canada-France Ecliptic Plane Survey and its high-latitude (HiLat) extension \citep[CFEPS;][]{pet11, pet17}. 
We refer to the combination of these three well-characterized surveys as OSSOS+.
Their sample of five NTs, together with their quantified survey characteristics, can provide a valuable constraint on the NT orbital distribution. 

In this study, we combine OSSOS+ and PS1 to assess the intrinsic eccentricity and inclination distribution of the Neptunian Trojans. 
Section~\ref{sec:sample} introduces the OSSOS+ and PS1 survey coverage and NT sample, with an analysis of the dynamical stability of the four of five OSSOS+ NTs in Section~\ref{sec:dynamics}. 
We use a survey simulator for both OSSOS+ and the PS1 survey to investigate the orbital distribution of the intrinsic population (Section~\ref{sec:simulation}), and discuss the evidence for two size-dependent components in the NT inclination distribution.
We discuss possible formation mechanisms for two-component (Section~\ref{sec:dis}). We conclude in Section~\ref{sec:sum}.

\section{The stable Neptunian Trojan sample from the OSSOS+ and PS1 surveys}
\label{sec:sample}

\begin{figure*}
\includegraphics[width = \textwidth]{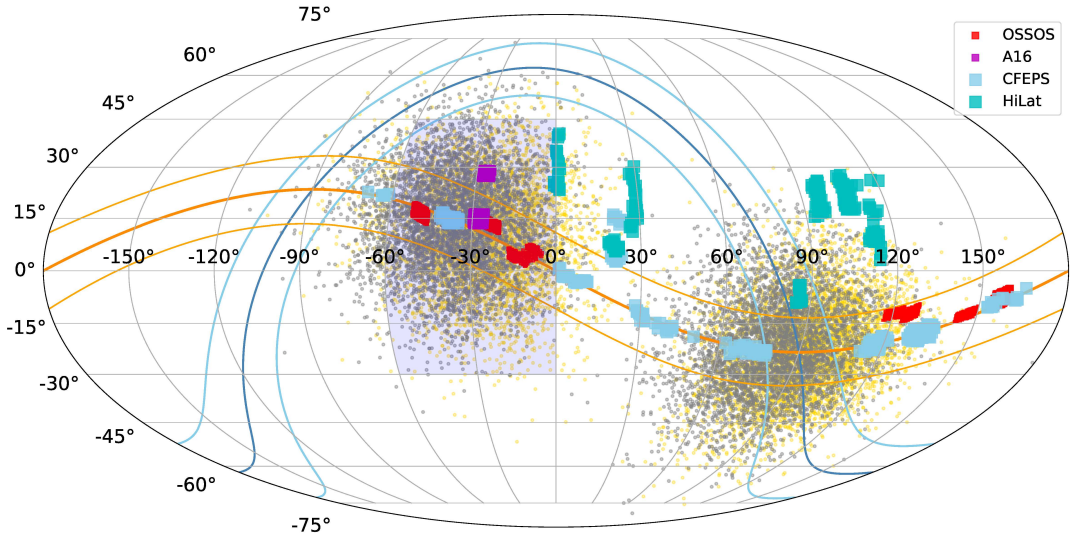}
\caption{Sky coverage of the surveys used in this analysis relative to the Neptunian Trojan population model developed in Sec~\ref{sec:simulation} (inclination width $\sigma_i = 15^{\circ}$, epoch = 2015 November 1 (grey dots) and epoch = 2010 November 1 (yellow dots) ). 
The L4 (left) and L5 (right) clumps are both shown relative to the ecliptic (thick orange line) and galactic (blue line) planes.
 Note that CFEPS ran most of a decade earlier than OSSOS, so they sample different parts of a moving population, despite the footprints partly overlapping.
The PS1 coverage is approximately indicated in two ways: the Solar System survey to $r \sim22.5$ encompasses the region within $\pm 12^{\circ}$ of the ecliptic plane (between two thin orange lines), while the 3$\pi$ survey to $r \sim21.5$ covered all the area north of Decl. $-30^{\circ}$. The pale blue shading area shows the effective coverage (avoid galactic plane) of PS1 3$\pi$ survey on the L4 clump.
} 
\label{fig:skyarea}
\end{figure*}

\subsection{NTs from OSSOS+}
\label{sec:ossos_sample}

The survey coverage of the well-characterized surveys that we term OSSOS+ neatly samples both sides and the middle of the on-sky distribution of the L4 Neptunian Trojans (Fig.~\ref{fig:skyarea}).
OSSOS had eight survey blocks of sky on or within $12\degr$ of the invariable plane, half of which were bracketing the margins of the L4 zone.
A16 had two survey regions, both within the L4 zone, with one on-plane and one at higher $\sim14\degr$ ecliptic latitudes, intended to constrain the L4 distribution.
Both surveys are deep, to 24th--25th magnitude in $r$; roughly consistent with H$_r = 10.5$ at 30~au, and H$_r$ is the r-band absolute magnitude.
CFEPS had fields distributed on or near the ecliptic across a wide range of longitudes, with its high ecliptic latitude (HiLat) extension sampling fields from latitudes of 12$\degr$ to 85$\degr$.
CFEPS/HiLat are shallower at 22nd--24th magnitude in $g$ and $r$.
Together, the deep and mostly low-latitude coverage of OSSOS+ is ideally placed to preferentially constrain the low-inclination L4 NTs.
However, as these are all surveys with well-quantified detection efficiencies, they also constrain the more rarely detected high-inclination NTs.  

There are five NTs from OSSOS+, all stable (Sec.~\ref{sec:gyr}) and from Neptune's L4 Lagrange region (Table~\ref{tab:orbits}).
OSSOS discovered four: 2015 VV$_{165}$, 2015 VW$_{165}$, 2015 VX$_{165}$, all in the on-plane 15BC block, and 2015 RW$_{277}$, in the on-plane 15BS block. These were among 843 outer Solar System discoveries by this wide-field imaging survey in 2013-2017, using the Canada-France-Hawaii Telescope (CFHT) \citep{ban18}.
The NT 2012 UV$_{177}$ was discovered by A16 in their higher-latitude survey region, also with CFHT.
No NTs were detected in CFEPS.

\subsection{NTs from Pan-STARRS1}

PS1 has two surveys in the period between May 2010 and 2014 that we consider: the shallower all-sky 3$\pi$ steradians survey, which has a limiting magnitude of $r_{\rm PS1}$-band limiting magnitude $\sim 21.5$, and the slightly deeper $w_{\rm PS1}$-band Solar System Survey with limiting magnitude $\sim 22.5$ (H$_r \sim 8$ at 30~au). 
Their sky coverage is across both L4 and L5 points (Fig.~\ref{fig:skyarea}), though like many other surveys, PS1's analysis does not provide moving-object detections in the Galactic plane, and thus is sensitive only to large-amplitude librators and not the core of the L5 region.  

In 2010-2014, PS1 detected seven NTs using the PS1 Outer Solar System Pipeline, including re-detection of two that were previously known, 2001 QR$_{322}$ and 2006 RJ$_{103}$ (Table~\ref{tab:orbits}). The detection algorithm was described in \citet{hol18}. Two of the seven, 2006 RJ$_{103}$ and the brightest detection, 2013 KY$_{18}$, were detected in the 3$\pi$ survey, with the other five NTs found in the Solar System Survey.
We exclude the only L5 NT in this PS1 sample, 2013 KY$_{18}$, from our population analysis (Sec.~\ref{sec:simulation}), as its dynamical lifetime of about a million years indicates it is only temporarily captured \citep{lin16}.

\section{Dynamical Characterization of the OSSOS Neptunian Trojans} 
\label{sec:dynamics}

\subsection{Orbits on 10 Myr timescale}
\label{sec:tenmyr}

Table~\ref{tab:orbits} lists the best-fit barycentric orbits for the four OSSOS NTs, and for the other currently known NTs, based on all available astrometry listed at the Minor Planet Center on April 2018. 
The best-fit barycentric orbit and the 1-sigma level uncertainties of the orbital parameters were obtained by the orbit fit routine of \citet{ber00}. 

The libration characteristics of all the known NTs were determined from 10 Myr integrations of the best-fit orbit and 250 clones generated from sampling the covariance matrix of the best-fit orbit. All integrations were performed using SWIFT \citep{lev94}, with the Sun and the four giant planets included as the only massive bodies in the system, and the mass of the terrestrial planets added into the Sun. 
We list in Table~\ref{tab:orbits} the libration center ($\phi_{center}$) and amplitude ($A_{\phi}$) for the resonant angle $\phi_{1:1} = \lambda_N - \lambda_T$, where $\lambda_N$ and  $\lambda_T$ are the mean longitude of Neptune and the Trojan, respectively.  
In order to be consistent with the results of \citet{ale16}, \citet{lin16} and \citet{par15}, the $A_{\phi}$ is presented as a half-peak amplitude, which is $\sqrt{2} \times \sigma_{\phi}$, where the $\sigma_{\phi}$ is the standard deviation of the resonant angle $\phi_{1:1}$.  

\begin{deluxetable*}{lllllllllllll}
\tabletypesize{\scriptsize}
\tablecaption{Barycentric orbits and resonant dynamics of the known Neptune Trojans \label{tab:orbits}}
\tablehead{Designation & a (au) & e & {\it i} (\degr) & $\phi_{center}$ (\degr) & $A_{\phi}$ (\degr) & $H_r$ & L & Stability & Discovery Survey}
\startdata
2001 QR$_{322}$  & 30.232 $\pm$ 0.001 & 0.02849 $\pm$ 0.00002 & 1.323 $\pm <$ 0.001 & $68.0^{+0.1}_{-0.1}$ & $26.4^{+0.2}_{-0.2}$ & 7.7 & L4 & M$^{*}$ & Deep Ecliptic Survey$^\P$\\ 
(385571) Otrera & 30.144 $\pm$ 0.003 & 0.0270 $\pm$ 0.0002 & 1.431 $\pm <$ 0.001 & $61.3^{+0.2}_{-0.1}$ & $10.8^{+0.8}_{-0.7}$ & 8.6 & L4 & S$^{*}$ & ST06 \\ 
2005 TN$_{53}$  & 30.125 $\pm$ 0.002 & 0.06584 $\pm$ 0.00003 & 25.001 $\pm <$ 0.001 & $59.16^{+0.09}_{-0.09}$ & $8.1^{+0.6}_{-0.6}$ & 8.8 & L4 & S$^{*}$ & ST06\\ 
(385695) Clete & 30.132 $\pm$ 0.001 & 0.05280 $\pm$ 0.00005 & 5.252 $\pm <$ 0.001 & $61.10^{+0.08}_{-0.06}$ & $8.5^{+0.4}_{-0.3}$ & 8.1 &  L4 & S$^{*}$ & ST06\\ 
2006 RJ$_{103}$  & 30.0393 $\pm$ 0.0009 & 0.03002 $\pm$ 0.00002 & 8.163 $\pm <$ 0.001 & $60.380^{+0.009}_{-0.009}$ & $5.01^{+0.07}_{-0.07}$ & 7.3 & L4 & S$^{*}$ & SDSS$^\P$\\
(527604) 2007 VL$_{305}$  & 30.027 $\pm$ 0.002 & 0.06331 $\pm$ 0.00003 & 28.117 $\pm <$ 0.001 & $61.00^{+0.05}_{-0.05}$ & $14.3^{+0.2}_{-0.2}$ & 7.7 & L4 & S$^{*}$ & SDSS\\ 
2008 LC$_{18}$  & 29.957 $\pm$ 0.003 & 0.08308 $\pm$ 0.00006 & 27.539 $\pm <$ 0.001 & $297.6^{+0.4}_{-0.4}$ & $17.2^{+1.2}_{-1.2}$ & 8.0 & L5 & M$^{*}$ & ST10 \\ 
2010 TS$_{191}$  & 30.008 $\pm$ 0.001 &  0.04586 $\pm$ 0.00003 & 6.563 $\pm <$ 0.001 & $61.83^{+0.02}_{-0.02}$ & $11.38^{+0.07}_{-0.07}$ & 7.8 & L4 & M$^{*}$ & PS1 \\ 
2010 TT$_{191}$  & 30.087 $\pm$ 0.004 & 0.07029 $\pm$ 0.00008 & 4.276 $\pm <$ 0.001 & $65.3^{+0.1}_{-0.1}$ & $19.5^{+0.3}_{-0.3}$ & 7.6 & L4 & M$^{*}$ & PS1 \\ 
2011 MH$_{102}$  & 30.111 $\pm$ 0.005 & 0.0806 $\pm$ 0.0005 & 29.377 $\pm $ 0.001 & $299.7^{+0.4}_{-0.4}$ & $9.9^{+2.0}_{-1.8}$ & 7.9 & L5 & S$^{*}$ & P13 \\ 
(530664) 2011 SO$_{277}$  & 30.162 $\pm$ 0.002 & 0.01187 $\pm$ 0.00003 & 9.639 $\pm <$ 0.001 & $63.9^{+0.2}_{-0.2}$ & $18.9^{+0.5}_{-0.4}$ & 7.4 & L4 & S$^{*}$ & PS1 \\
(530930) 2011 WG$_{157}$  & 30.030 $\pm$ 0.003 & 0.02791 $\pm$ 0.00007 & 22.299 $\pm <$ 0.001 & $61.38^{+0.02}_{-0.01}$ & $15.69^{+0.05}_{-0.04}$ & 6.9 & L4 & S$^{*}$ & PS1 (3$\pi$) \\ 
2012 UD$_{185}$ & 30.201 $\pm$ 0.002 & 0.04406 $\pm$ 0.00004 & 28.299 $\pm <$ 0.001 & ? & ? & 7.4 & L4 & ? & PS1 (IfA) \\
2012 UV$_{177}$  & 30.024 $\pm$ 0.004 & 0.0723 $\pm$ 0.0008 & 20.833 $\pm <$ 0.001 & $60.7^{+0.2}_{-0.2}$ & $9.8^{+0.9}_{-0.7}$ & 9.0 & L4 & S$^{+,*}$ & A16 \\ 
2013 KY$_{18}$  & 30.149 $\pm$ 0.005 & 0.123 $\pm$ 0.002 & 6.658 $\pm <$ 0.001 & $293.4^{+1.3}_{-2.0}$ & $20.6^{+4.1}_{-2.6}$ & 6.6 & L5 & T$^{*}$ & PS1 \\ 
2013 VX$_{30}$  & 30.0876 $\pm$ 0.0006 & 0.08374 $\pm$ 0.00002 & 31.259 $\pm <$ 0.001 & $59.09^{+0.02}_{-0.02}$ & $5.1^{+0.2}_{-0.2}$ & 8.1 & L4 & S$^{\clubsuit}$ & Dark Energy Survey\\ 
2014 QO$_{441}$  & 30.1019 $\pm$ 0.0007 & 0.10528 $\pm$ 0.00004 & 18.831 $\pm <$ 0.001 & $61.71^{+0.05}_{-0.04}$ & $10.3^{+0.2}_{-0.2}$ & 8.1 & L4 & S$^{*}$ & Dark Energy Survey\\ 
2014 QP$_{441}$  & 30.074 $\pm$ 0.003 & 0.0650 $\pm$ 0.0008 & 19.403 $\pm <$ 0.001 & $59.46^{+0.07}_{-0.04}$ & $2.0^{+0.8}_{-0.5}$ & 9.1 & L4 & S$^{*}$ & Dark Energy Survey\\ 
2014 UU$_{240}$  & 30.057 $\pm$ 0.001 & 0.0484 $\pm$ 0.0001 & 35.744 $\pm <$ 0.001 & $57.1^{+0.3}_{-0.1}$ & $2.6^{+0.3}_{-0.4}$ & 8.1 & L4 & S$^{\clubsuit}$ & Dark Energy Survey\\ 
2015 RW$_{277}$ & 30.013 $\pm$ 0.005  & 0.077 $\pm$ 0.001 & 30.826 $\pm$ 0.001 & $65.8^{+0.9}_{-1.1}$ & $27.8^{+2.6}_{-4.1}$ & 9.92 & L4 & M$^{\dag}$ & OSSOS\\ 
2015 VV$_{165}$ &  30.120 $\pm$ 0.001 & 0.0850 $\pm$ 0.0001 & 16.855 $\pm <$ 0.001 & $64.5^{+0.9}_{-0.6}$ &  $20.5^{+2.0}_{-1.3} $  & 8.83 & L4 & M$^{\dag}$ & OSSOS\\
2015 VW$_{165}$ & 30.102 $\pm$ 0.001 & 0.0511 $\pm$ 0.0002 & 4.998 $\pm <$ 0.001 & $63.3^{+0.6}_{-0.4}$& $16.3^{+1.7}_{-1.0}$  & 8.03 & L4 & S$^{\dag}$ & OSSOS \\
2015 VX$_{165}$ & 30.073 $\pm$ 0.001 & 0.07522 $\pm$ 0.00003 & 17.140 $\pm$ 0.001 & $61.63^{+0.02}_{-0.06}$ & $12.6^{+0.1}_{-0.2} $ & 8.85 & L4 & S$^{\dag}$ & OSSOS\\
\enddata
\tablenotetext{}{Notes --- a, e and i were fitted from all available observations, with an epoch of the first day of observation. \\ $\phi_{center}$ and $A_{\phi}$ were found by 10 Myr integration (Sec.~\ref{sec:tenmyr}). \\ $H_r$ of the non-OSSOS NTs were computed from the Minor Planet Center's reported $H_V$, assuming that V-r = 0.2.\\ L --- Lagrange point of 1:1 resonance.\\ Stability --- S: dynamically stable for $\geq1$ Gyr; M: metastable for most of a Gyr; T: transiently in 1:1 resonance with lifetime on the order of Myr. Reference: $^{\dag}$This work . $^{\clubsuit}$\citet{lin19}. $^{+}$\citet{ale16}. $^{*}$\citet{wu19}.}

\tablenotetext{}{Discovery surveys --- Deep Ecliptic Survey: \citet{ell05}. ST06: \citet{she06}.  ST10: \citet{she10b}. P13: \citet{par13}. PS1: \citet{lin16}. PS1 (IfA): Only recently listed at the MPC in \href{http://www.minorplanetcenter.net/db_search/show_object?object_id=2012+UD185}{MPS 990734, MPS 990735}. A16: \citet{ale16}. Dark Energy Survey: \citet{ger16,lin19}. OSSOS: \citet{ban16, ban18}. $\P$: Also detected by PS1.}
\end{deluxetable*}

Remarkably, the five NTs from the OSSOS+ NT sample mostly have {\it high} orbital inclinations. Four have inclination $\gtrsim 17\degr$, three of which were detected in OSSOS survey regions centered on the invariable plane. Only one, 2015 VW$_{165}$, has a low inclination of $\sim5\degr$. Considering the predominantly low ecliptic latitudes of the two surveys, the generally high inclinations of these NTs are surprising.
We consider the implications for a wide NT inclination distribution in Section~\ref{sec:simulation}.

\subsection{Dynamical Stability over 1 Gyr}
\label{sec:gyr}

The long-term dynamnical stability of each of the known NTs is indicated in Table~\ref{tab:orbits}; the stability of most of these NT has been determined in previous works.
To understand the long-term dynamical stability of the four OSSOS NTs, we integrated 1000 clones of each object for 1 Gyr. The clones were sampled from the six-parameter covariance matrix of the best-fit orbit and integrated with the same set-up as for the 10 Myr computations (Sec.~\ref{sec:tenmyr}). 
We calculated the mean lifetime $\tau_0$ ($\tau_0 = 1/\lambda$, $\lambda$ is the exponential decay constant) of the clones by fitting an single exponential decay function to the number of clones remaining in the 1:1 resonance, as a function of the integration time.  
We consider that once a clone leaves the resonance, it is lost (not remaining) regardless of a potential return into the resonance.
The results are summarized in Table~\ref{tab2}, and we consider each NT in term below.

\begin{table}
\caption{Dynamical lifetimes of the OSSOS Neptunian Trojans \label{tab2}}
\begin{tabular}{lcc}
\hline
Trojans &  $\tau_0$ (Gyr) $^a$ & $\tau_1$ (Gyr) $^a$\\
\hline
\hline

2015 RW$_{277}$ & 1.4 & 0.11 \\ 
2015 VV$_{165}$ & 0.65 & -- \\
2015 VW$_{165}$ & $> 4.5$ $^b$ & -- \\
2015 VX$_{165}$ & 2.81 & --\\
\hline
\end{tabular}
\\
$^a$ $\tau_0$ and $\tau_1$ are the mean lifetimes of two-phase exponential decay.
$^b$ None of the 1000 clones were lost during the 1 Gyr integration.\\
\end{table}

\subsubsection{Highly stable: 2015 VW$_{165}$ and 2015 VX$_{165}$}
The clones of 2015 VW$_{165}$ are extremely stable: none of its thousand clones were lost during the 1 Gyr integration.
The clones of 2015 VX$_{165}$ also show good stability with $\tau_0 \sim 2.81$ Gyr; less than a third of its thousand clones were lost during the 1 Gyr integration. The fact that most of the allowable orbit-fit parameter space for these two NTs is long-term stable supports the conclusion that they were likely captured into the NT population early in the Solar System's history.

\subsubsection{Complex lifetimes of 2015 VV$_{165}$ and 2015 RW$_{277}$}

\begin{figure}
\includegraphics[width = .5\textwidth]{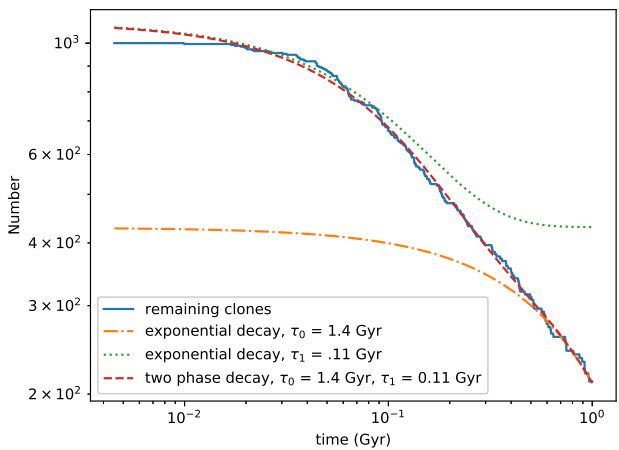}
\caption{The retention over a Gyr in the 1:1 resonance of a thousand clones of the Neptune Trojan 2015 RW$_{277}$'s orbit, sampled from within its covariance matrix of orbital uncertainties. The best-fit two-phase exponential decay curve shows in red dashed line, which is the superposition of two exponential decay curves, the orange dash-dotted line with mean lifetimes $\tau_0$ = 1.4 Gyr and the green dotted line with $\tau_1$ = 0.11 Gyr. This metastable behaviour has been seen previously in the NTs, by 2001 QR$_{322}$.
\label{fig:metastable}}
\end{figure}

The cases of 2015 VV$_{165}$ and 2015 RW$_{277}$ are more complicated. 
The clones of 2015 VV$_{165}$ show meta-stability. Only about a quarter of the thousand clones survived for 1 Gyr, and the mean lifetime of the clones is only 0.65 Gyr. 
2015 RW$_{277}$'s clones show diverse behaviour (Figure~\ref{fig:metastable}).  
About half of 2015 RW$_{277}$'s clones have $\tau_0 > 1$ Gyr; the other half have a shorter lifetime of $\tau_1 \sim 130$ Myr. 
Their decay is only well fit by two exponential decay functions, each with a different mean lifetime, $\tau_0$ = 1.4 Gyr, and $\tau_1$ = 0.11 Gyr. 
The lifetime of 2015 RW$_{277}$'s clones are correlated with their orbital elements, so the diversity of the decay rates is likely due to this object's observational uncertainties; their dynamical stability will be worth to re-assessing if additional observations are acquired.

Meta-stability has been seen previously in the NTs: for instance, 2001 QR$_{322}$ and 2008 LC$_{18}$ showed similar behaviors and lifetime distributions \citep{bra04b, hor10, hor12b, hor16, lyk11, zho11} and presented very similar decay curves \citep{hor16}. 
Even though the clones of 2015 VW$_{165}$ and 2015 RW$_{277}$ are dynamically metastable, they still have much longer lifetimes than a temporarily captured NT, which typically have only Myr-scale lifetimes \citep{gua12, hor06, hor12b, ale13}. 
We do not consider it plausible that these meta-stable NTs are captured objects on unusually long Gyr timescales; \citet{ale13} found a mean lifetime of just 78 kyr and a maximum of 18.2 Myr in their simulations of temporary NTs. 
Given this, it is likely that future refinements to their orbits will reveal that they are, in-fact, long-term stable. Thus, we consider 2015 VV$_{165}$ and 2015 RW$_{277}$ as in the stable population, rather than as temporary captures.

\section{Survey constraints on the intrinsic Neptune Trojan orbital distribution and population}
\label{sec:simulation}


To place constraints on the intrinsic population of NTs, we use a survey simulator to apply the surveys' quantified detection biases (to the degree available) to an NT population model. We adjust the model's parameters until, using the two-sample Anderson-Darling test (AD test, which is more sensitive to the tails of the distribution than the Kolmogorov-Smirnov test), the model produces simulated detections that match the observed number and orbital element distribution of our observed NT sample.

The OSSOS survey simulator \citep{ban16, law18a} has been used in previous works to model orbital distributions and population estimates \citep[e.g.][]{kav09, pet11, vol16, vol18, sha17}. 
The quantified detection efficiencies have been measured for all the OSSOS+ surveys and incorporated into the survey simulator.

For the PS1 sample, a quantified survey efficiency is not yet available.
PS1 is a highly multiplexed survey, and the chance of detecting an object increases with the number of times that it could have been observed. Therefore, we adopted the observing selection function from \citet{lin16} for the PS1 Solar System Survey, which based on the number of visits and worked well to estimate the NT detectability.  It can be summarized as the total sum of probability mass functions of binomial distribution:
\begin{equation}
\label{ps1_eff}
f_{eff}(n) = \sum\limits_{i=10}^n\binom{n}{i}0.35^i\times(1-0.35)^{n-i}.
\end{equation}
Here, $n$ is the total number of exposures in a specific survey region, $i$ is the minimal number of detections required for finding an object in \citet{lin16}, which is 10, and 0.35 is the 50$\%$ of detectability of r=22.5 object for Solar system survey, or 21.5 objects for 3$\pi$ survey times the filling factor of 70$\%$. 

Thus, the detectability of PS1 is a function of number of exposures. For the Solar system survey, the typical number of exposures is between 10 to 60 and vary with sky positions.
For the $3\pi$ survey, because there were so many repeated visits to various areas of the sky, typically larger than 100, the equation~\ref{ps1_eff} would eventually close to 1 for a magnitude 21.5 object. However, since the galactic plane complicates the detection efficiency due to a much more crowded stellar field, and the assumption of 100$\%$ efficiency presumably would not be sufficient. Therefore, We define a effective $3\pi$ survey coverage field of the sky around L4 ($0^\circ <$ RA $< 60^\circ$ and $-30^\circ <$ Decl. $< 45^\circ$, see Figure~\ref{fig:skyarea}).

We generated models of the NT orbital distribution with a fixed semi-major axis (a=30.1 au), a Rayleigh distribution of eccentricity, a Rayleigh distribution of the libration amplitude and a sin(i) $\times$ Gaussian distribution of inclination. Each of these distributions has just a single free parameter, to avoid the over-fitting of a small number of observations. The Rayleigh distribution is given by:
\begin{equation}\label{reyleigh}
p(x; \sigma) \propto x e^{-\frac{1}{2}(x/\sigma)^2},  x \geq 0.
\end{equation}
Here $\sigma$ is the width of the Rayleigh distribution, and we use it to be the width of the eccentricity distribution, $\sigma_e$, or the width of the libration amplitude distribution, $\sigma_{A_\phi}$. The sin(i) $\times$ Gaussian distribution is an Rayleigh distribution with sin(i) instead $x$ in equation~\ref{reyleigh}, which is:

\begin{equation}
p(\textrm{sin(i)}; \sigma_i) \propto \textrm{sin(i)} e^{-\frac{1}{2}(\textrm{sin(i)}/\sigma_{i})^2},  \textrm{sin(i)} \geq 0.
\end{equation}
Similar models were used in \citet{ale16} and \citet{lin16}. 

The cumulative absolute magnitude distribution function is given by a single power-law distribution:
\begin{equation}
\Sigma(\textrm{H}) \propto 10^{\alpha \textrm{H}},
\end{equation}
Here H is the absolute magnitude.
We tested two different absolute magnitude distributions applied to the NT models, with power-law index $\alpha = 0.9$, the bright end slope of dynamical excited TNOs \citep{fra14,law18b}, and $\alpha = 0.65$, the average slope of TNOs \citep{ber04}. 
We expect the real orbital/absolute magnitude distributions to be more complicated \citep{vol16, law18b}. However, based on the results of Figure~\ref{phi_prob} and later, since there is little difference between the different models, the single parameter/slope is fine for such a small number of detections.

To constrain acceptable models for the intrinsic orbital element distribution (the libration amplitudes, the eccentricity or the inclination), we ran the survey simulations and calculated the AD probabilities of the relevant orbital element distributions for the simulated and real observations, to test which model is most consistent with the OSSOS+ and PS1 NT detections. 
We simulate the two surveys entirely independently to permit cross-checks.
For each model, we simulated 2000 NT detections. 
The AD statistics were calculated by selecting 1000 sub-samples from the 2000 simulated NT detections, with each sub-sample containing the same number of NTs in the observed samples (i.e. 6 observed NTs for PS1 and 5 for OSSOS+). 
Finally, we calculated the AD statistics of the sub-samples' relevant orbital element distribution, versus those of the whole simulated samples, to understand the null distributions of each NT model. 
We estimated the rejectability of each model by comparing the AD statistics of the observed samples with the null distributions. 

\subsection{Libration Amplitude Width} \label{sec:amp}

We test the acceptable range of $\sigma_{A_\phi}$ for both OSSOS+ and PS1. 
The inclination width of the NT model for OSSOS+ and PS1 survey simulations were set as 22\degr\ and 6\degr, respectively (see Section~\ref{sec:inc}). The eccentricity width were set as 0.07 and 0.27, respectively (see Section~\ref{sec:ecc}).
With the inclination width and eccentricity width fixed, we generated models with $\sigma_{A_\phi}$ set from 5\degr\ to 40\degr\ with 1\degr\ steps.
We widen the libration amplitude from \citet{par15}'s best suggested $\sigma_{A_\phi} = 10\degr$, as most of the NTs in our sample have $A_\phi > 10\degr$ (Table~\ref{tab:orbits}).

\begin{figure}
\includegraphics[width = .5\textwidth]{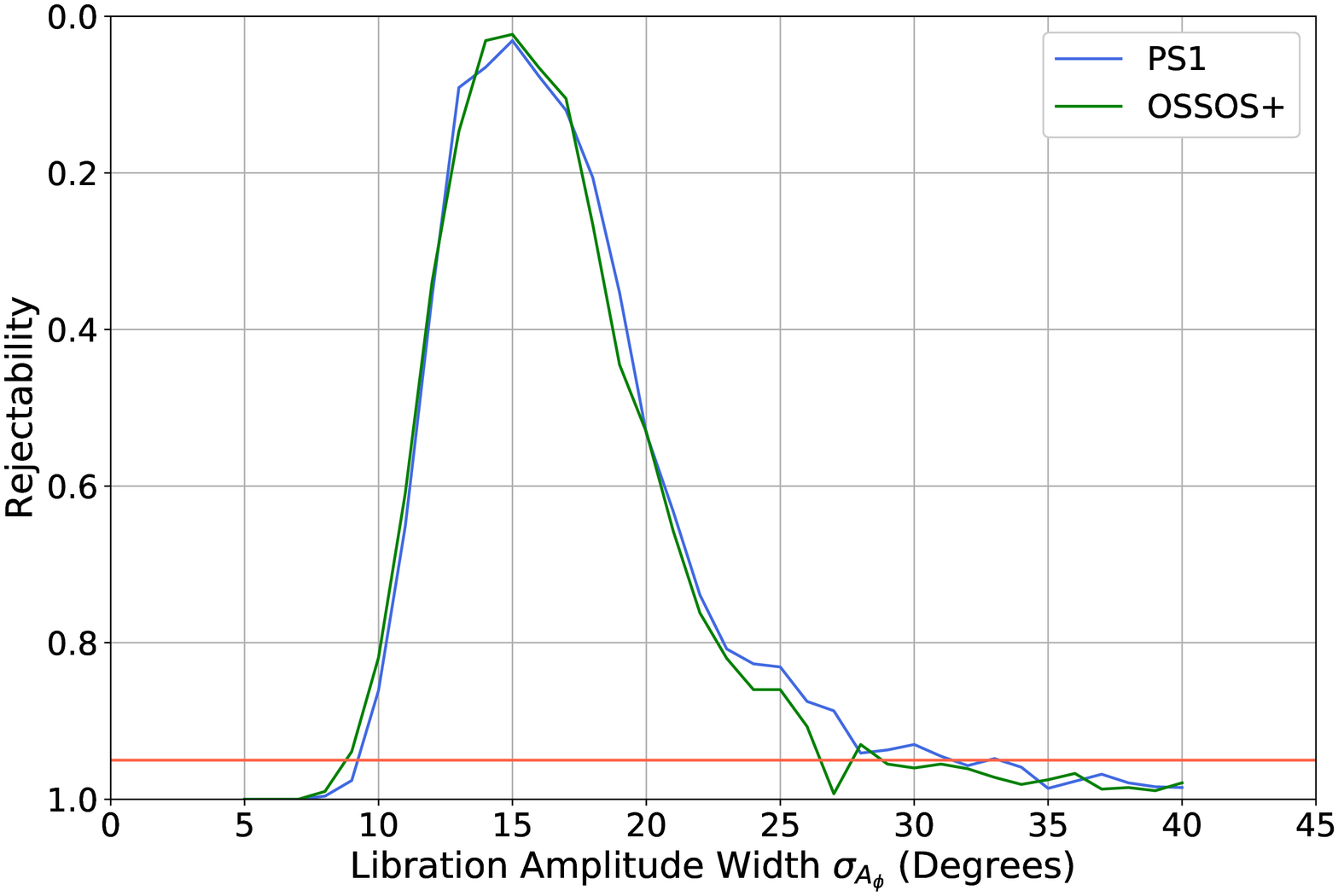}
\caption{The bootstrapped AD rejectability of various values of the libration amplitude width $\sigma_{A_\phi}$. Values below the red line are rejectable at greater than 95$\%$ confidence.}
\label{phi_prob}
\end{figure}

We find, as shown in Figure~\ref{phi_prob}, if the libration amplitude distribution of NTs follow a Rayleigh distribution, the libration amplitude width, $\sigma_{A_\phi}$, must be greater than 5\degr. Moreover, PS1 and OSSOS+ have results consistent with each other. Figure~\ref{phi_prob} shows that $\sigma_{A_\phi} = 15^{\circ}$ has the lowest rejectability to both samples, though $\sigma_{A_\phi} = 10\degr$ as suggested by \citet{par15} is not rejectable. 

\subsection{Eccentricity Width} 
\label{sec:ecc}

\begin{figure}
\includegraphics[width = .5\textwidth]{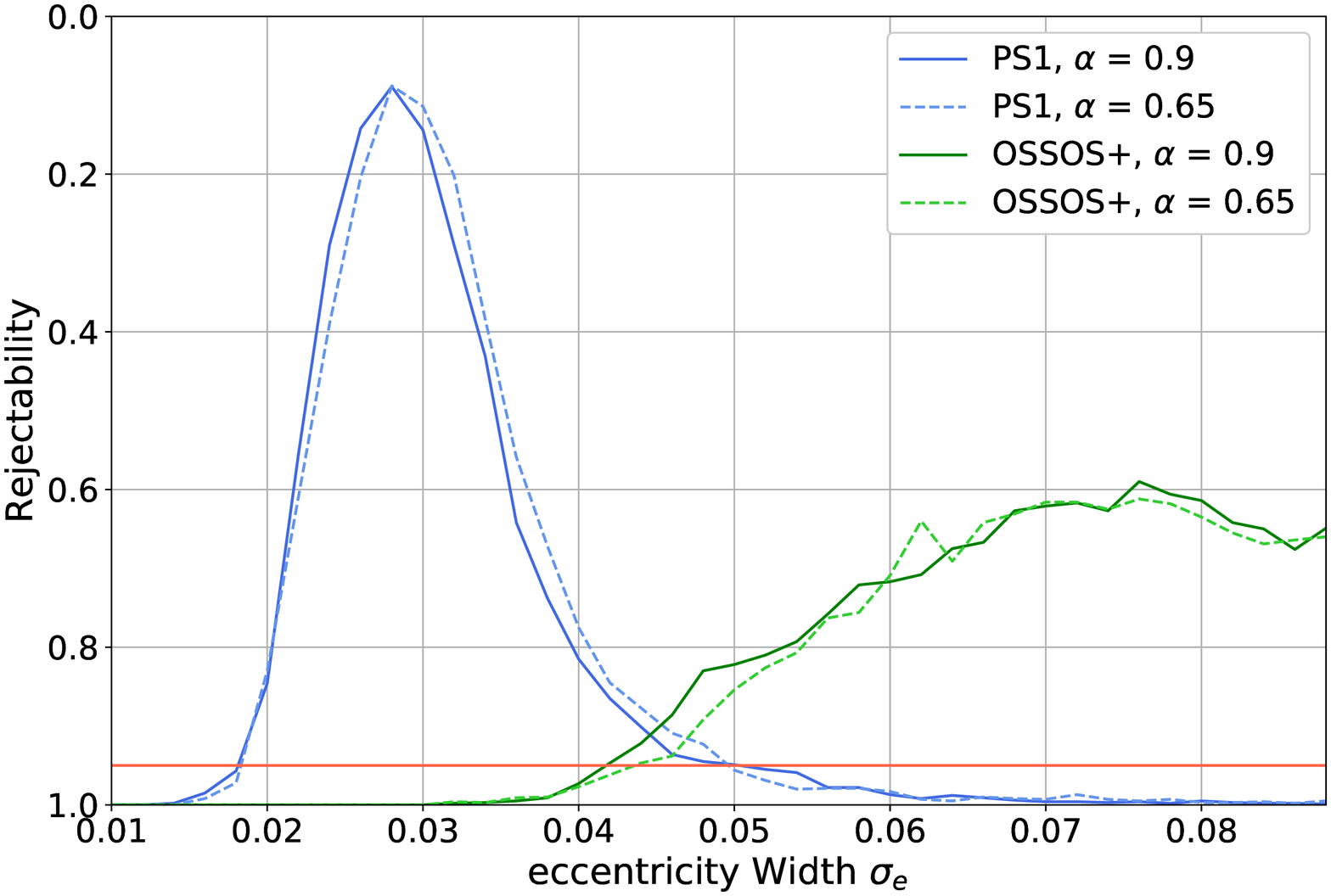}
\caption{The bootstrapped AD rejectability of various values of the eccentricity width $\sigma_e$ of the Neptune Trojan orbital distribution.
Values below the red line are rejectable at greater than $95\%$ confidence. The non-rejectable overlap area between the OSSOS+ and PS1 constraints on the eccentricity width is about 0.045. Varying the size distribution $\alpha$ makes little difference to the results.  \label{fig_ecc}}
\end{figure}

We generated models with the eccentricity width ($\sigma_e$) of the Rayleigh distribution varied from 0.01 to 0.09, and constrain acceptable models in the same way as with the simulations of $\sigma_{A_\phi}$ (Sec.~\ref{sec:amp}), but while varying the eccentricity. Fig~\ref{fig_ecc} shows that the rejectability of various eccentricity models is startlingly different between the two surveys.  The result suggests that if the eccentricity distribution follows Rayleigh distribution, the PS1 sample favors a generally smaller and narrower eccentricity distribution than the distribution of the OSSOS+ sample. Using a smaller slope ($\alpha = 0.65$) of absolute magnitude distribution function does not significantly affect the acceptable range of eccentricity width.

\subsection{Inclination Width} 
\label{sec:inc}

With the distribution of all orbital elements fixed to their least rejectable values except inclination, we generated 14 different models with the inclination width ($\sigma_i$) of the sin(i) $\times$ Gaussian distribution varied from 4 to 30\degr\ with a 2\degr\ step.
The reference plane is the invariable plane of the Solar System \citep{sou12}.
We constrain acceptable models in the same way as with the simulations of $\sigma_{A_\phi}$ (Section~\ref{sec:amp}) and $\sigma_{e}$ (Section~\ref{sec:ecc}), but while varying the inclination. 

\begin{figure}
\includegraphics[width = .5\textwidth]{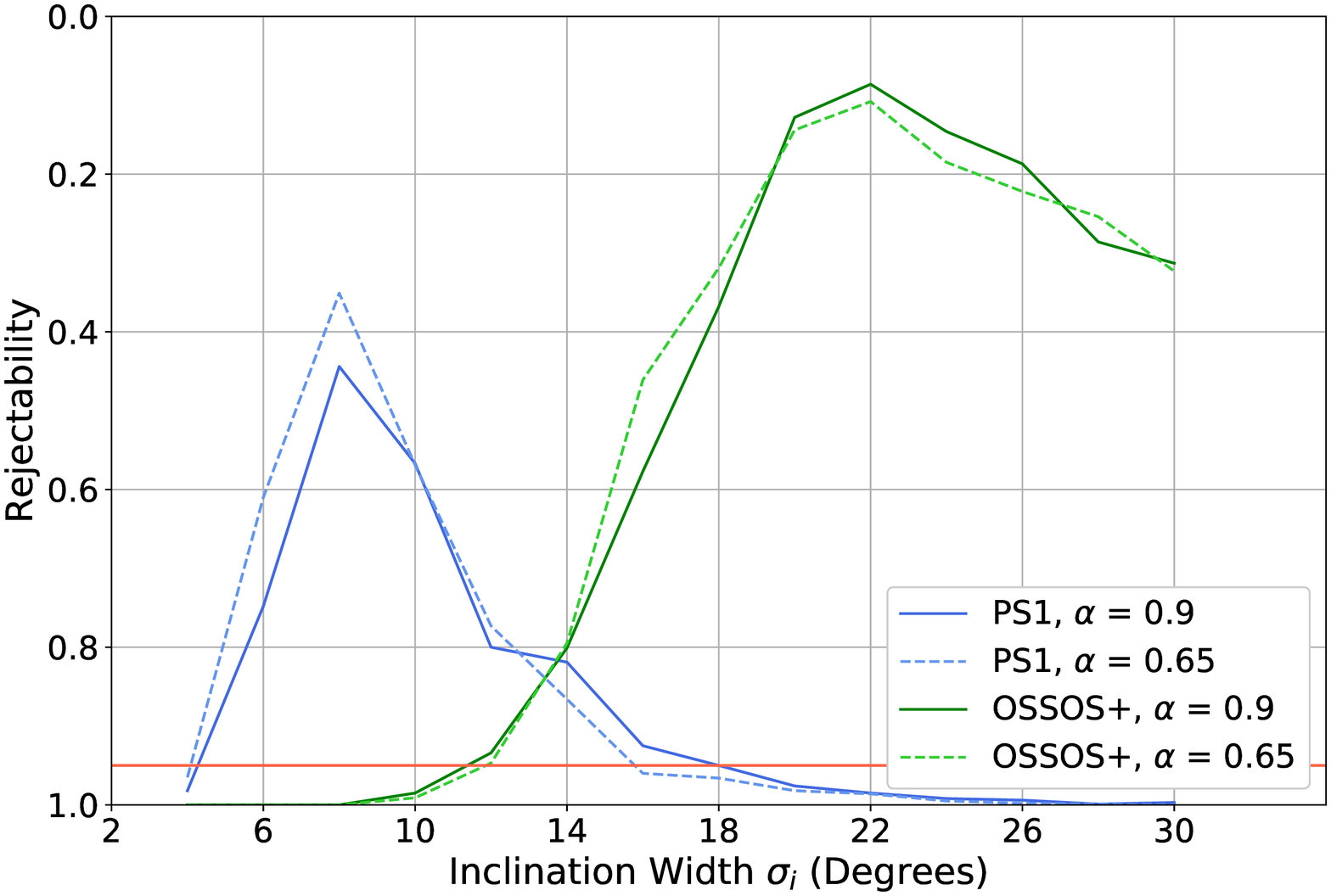}
\caption{The bootstrapped AD rejectability of various values of the inclination width $\sigma_i$ of the Neptune Trojan orbital distribution.
Values below the red line are rejectable at greater than $95\%$ confidence. The non-rejectable overlap area between the OSSOS+ and PS1 constraints on the inclination width is 12--16$^{\circ}$. Varying the size distribution $\alpha$ makes little difference to the results.  \label{fig:inc}}
\end{figure}

Fig~\ref{fig:inc} shows that the rejectability of various inclination models is startlingly different between the two surveys. 
The PS1 NT sample suggests that if the stable NT population has a sin(i) $\times$ Gaussian inclination distribution, the acceptable $\sigma_i$ range should be $\lesssim 16\degr$, while the OSSOS+ sample requires a higher inclination distribution with $\sigma_i \gtrsim 12\degr$. Similar to the $\sigma_e$ test, the two different $H$ distributions that we chose do not affect the result.


For the mutually acceptable range of $\sigma_i$, we consider the overlapping region where both surveys are not rejectable at the $95\%$ level. 
This intrinsic inclination distribution has a sin(i)~$\times$~Gaussian distribution with $\sigma_i$ of 12--16\degr. 
This result is very similar to the measured inclination widths for other resonant populations: 
Jovian Trojans ($\sigma_i \sim 14^\circ$, \citet{par15}), 
plutinos ($\sigma_i \sim 12^\circ$ in \citet{vol16} and $\sim 14^\circ$ in \citet{ale16}), 
5:2 resonators ($\sigma_i \sim 11^\circ$ in \citet{vol16} and $14^\circ$ in \citet{gla12}) and the 
5:3 resonators ($\sigma_i \sim 11^\circ$, \citet{gla12}). 

However, the acceptable $\sigma_i$ and $\sigma_e$ intervals of the two surveys are very different, and there is no easily evident detection bias in PS1 against finding high-inclination Trojans, considering PS1 has detected a high-inclination/retrogade object (471325) 2011 KT$_{19}$ \citep{niku2016}. Moreover, PS1 and OSSOS+ have different limiting magnitude (H$_r \sim 8$ and 10.5 at 30au, respectively) it is worthwhile to consider if the data are better represented by an alternative, size-Dependent model.

\subsection{A Bimodal and Size-Dependent NT Eccentricity/Inclination Distribution}
\label{sec:twodistrib}

Bimodal and size-dependent inclination distributions are known elsewhere in the trans-Neptunian populations. The classical Kuiper belt has a well-known bimodal inclination distribution \citep{bro01, lev01}. 
The inclinations of classical KBOs can be fitted by two sin(i) $\times$ Gaussian distributions with different $\sigma_i$ \citep{bro01, gul10}.
At least two \citep[`cold' and `hot';][]{ell05} or more \citep[`kernel', `stirred', and `hot';][]{pet11} populations exist in the classical Kuiper belt.
The luminosity function of each is distinct \citep{fra14}.

We test if the NT population could have a bimodal and size-dependent inclination distribution.
In this scenario, the physically larger ($ H_r < 8$) NTs mostly occupy dynamically cold, low eccentricity and low inclination orbital phase space (i $< 10^\circ$), and the physically smaller ($ H_r > 8$) NTs exist in dynamically hotter, high eccentricity and high inclination phase space.
To test this scenario, we mixed a `cold' component and an `hot' component in our NT model. We found that to match the real detections of PS1 and OSSOS+, the `cold' component needs a shallower H distribution with $\alpha \sim 0.2$, a cutoff at $ H_r \sim 8$, a eccentricity distribution with $\sigma_e \sim 0.02$, and an inclination distribution with $\sigma_i \sim 6^\circ$, which is similar to the 2:1 resonant population ($\sim 7^\circ$ in \citet{gla12} and $\sigma_i \sim 6^\circ$ in \citet{che19}). We set the the cutoff at $ H_r \sim 8$ based on the fact of only one $ H_r < 8$ NT has inclination $> 10^\circ$. Moreover, this cutoff avoids OSSOS+ detecting too many synthetic low inclination NTs. 

The `hot' component does not need any cutoff in H distribution. We set a `divot' absolute magnitude distribution, which have a bright-end slope and transitioning to a faint-end slope with a differential number contrast (see \citet{law18b} for detail description), $\sigma_e \sim 0.05$ and larger $\sigma_i \sim 18^\circ$, similar to the 5:1 resonant population ($\sigma_i \sim 22^\circ$, \citet{pik15}). We note that the selection of the two power laws `divot' or single power law absolute magnitude distribution has very little effect on the results of orbital distribution. The reason we choose `divot' here is based on the fact of that NT population lack the intermediate-sized members \citep{she10}, and it is reasonable for the total population estimation (see section~\ref{sec:pop}).

However, does the more complex two-component model represent the underlying NT population better than the simple single component model? We therefore evaluate the model preference using the Bayesian model comparison \citep{kass95}. To do so, we calculate the Bayes factor, $K$, between two models:

\begin{equation}
\label{eq:bayes_factor}
\begin{aligned}
K & = \frac{Pr(D|M_1)}{Pr(D|M_2)} \\
& = \frac{\int Pr(\theta_1|M_1) Pr(D|\theta_1, M_1)d\theta_1}{\int Pr(\theta_2|M_2) Pr(D|\theta_2, M_2)d\theta_2} \\
& = \frac{Pr(M_1|D)}{Pr(M_2|D)}\frac{Pr(M_2)}{Pr(M_1)}.
\end{aligned}
\end{equation}
Here $D$ is observed data, $M_1$ and $M_2$ are the plausibility of the two different models, and $\theta_1,  \theta_2$ are the model parameters of $M_1$ and $M_2$, respectively. If we propose the model priors are equal, $Pr(M_2) = Pr(M_1)$, the Bayes factor is the ratio of the posterior probabilities of $M_1$ and $M_2$.

To estimate the posterior probabilities, we use Approximate Bayesian computation (ABC) rejection algorithm. In this schema, a set of parameters $\theta$ that define the properties of model $M$ are first sampled from a prior distribution. From this sampled parameter set $\theta$, a data set $D_{sim}$ is simulated under the model $M$ specified by $\theta$. A similarity metric $\rho(D_{sim}, D_{obs})$ represents the similarity between simulated data set $D_{sim}$ and observed data $D_{obs}$. If the simulated data set $D_{sim}$ is too different from the observed data $D_{obs}$, the sampled parameters $\theta$ can be rejected. Therefore, we can set a cut off value $\epsilon$, and the acceptance criterion in ABC rejection algorithm is:
\begin{equation}
\rho(D_{sim}, D_{obs}) \leq \epsilon.
\end{equation}
We preform the ABC rejection and the Bayesian model comparison by the following steps:      
\begin{enumerate}
    \item Pick $\theta_p$ ($\sigma_e-cold$, $\sigma_e-hot$, $\sigma_i-cold$ $\sigma_i-hot$ for two components model, and $\sigma_e$, $\sigma_i$ for single component model) from the prior distribution (Table~\ref{tab:para}) to build Neptune Trojan model $M_p$. 
    \item Run survey simulation to generate 2000 simulated detections $D_{sim_p}$ based on the above model. 
    \item Calculate similarity metric $\rho(D_{sim_p}, D_{obs})$. Use two-sample AD statistics to determine a similarity $\rho$. For two dimensions ($\sigma_e$ and $\sigma_i$), the similarity metric $\rho$ is the sum of the two one-dimensional AD statistics divided by two.
    \item The cut off value $\epsilon$ is set dynamically. Similar to the bootstrapped AD statistics performed in section~\ref{sec:simulation}, randomly drew 6 PS1 and 5 OSSOS+ simulated detections for hundred times to determine the null distribution. Set a dynamical $\epsilon$ at top $5\%$ level of null distribution. Therefore, if a trial has $\rho$ smaller than $\epsilon$, keep $\theta_i$ as a successful trial.
    \item Repeat steps 1—4 until a sufficient number of trials have been successful.
    \item Finally, the ratio of the acceptance rates of two models is approximately the Bayes factor.
\end{enumerate}

We noted that the Bayes factors could be sensitive to the prior distribution of parameters. We use wider prior distributions on the two-component model than the single-component model (Table~\ref{tab:para}), and such prior distributions selection is a benefit for the acceptance rate of the single-component model; nevertheless, we find a Bayes factor $K \sim 7$,  which is substantial evidence that the two-component model is the preferred model for modeling NT population \citep{kass95}.

\begin{table}
\caption{Prior distributions\label{tab:para}}
\scalebox{0.8}{
\begin{tabular}{lccc}
\hline
Parameter & Distributions & single-component & two-component\\
\hline
\hline
$\sigma_{e-cold}$ &uniform & -- & 0.015 -- 0.045 \\ 
$\sigma_{e(-hot)}$&uniform & 0.04 -- 0.06 & 0.045 -- 0.07 \\
$\sigma_{i-cold}$&uniform & -- & $4^{\circ}$ -- $14^{\circ}$ \\ 
$\sigma_{i(-hot)}$&uniform & $10^{\circ}$ -- $18^{\circ}$ & $14^{\circ}$ -- $26^{\circ}$ \\
\hline
\end{tabular}}
\end{table}

\begin{figure}
\includegraphics[width = \columnwidth]{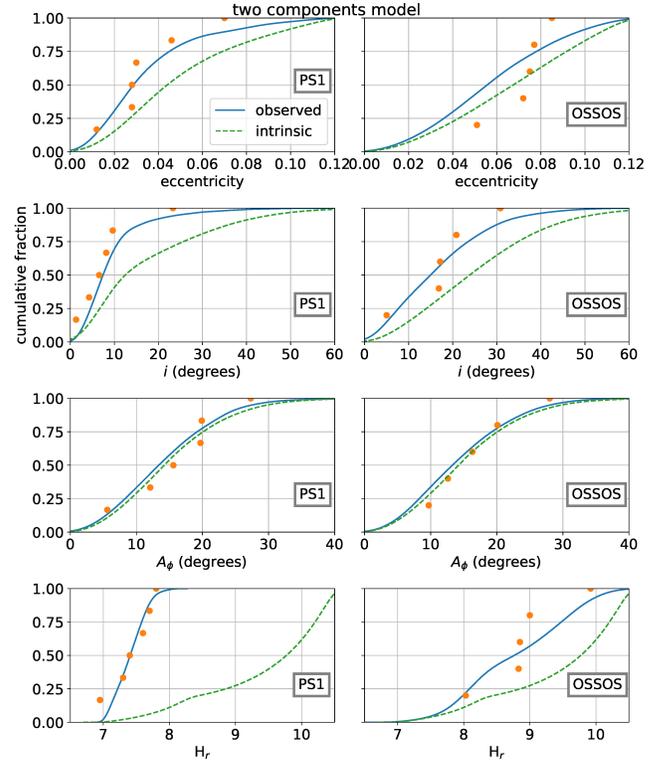}
\caption{The cumulative distributions of the eccentricity (top), inclination (2nd row), libration amplitude $\sigma_{A_\phi}$ (3rd row), and absolute magnitude H$_r$ (bottom) for the PS1 (left) and OSSOS+ (right) NT detections. The solid lines are the distribution of synthetic detections generated by the survey simulator. The dash lines are the intrinsic distributions of NT model. The dots show the real defections. Since the eccentricity and inclination distributions are absolute magnitude dependence, the top and 2nd row left panels show the intrinsic distributions of the NT model with H $\leq 8$.} 
\label{sim_vs_obs}
\end{figure}

Fig~\ref{sim_vs_obs} shows the cumulative distributions of the synthetic and real detections with our nominal values of $\sigma_{e_\phi}$, $\sigma_{i_\phi}$ and $\sigma_{A_\phi}$ and $\sigma_i$. 
The libration amplitude Rayleigh distribution with a width $\sigma_{A_\phi} = 12^{\circ}$ agrees well with both the PS1 and OSSOS+ detections.
The inclination distributions of the PS1 and OSSOS+ detections are well-matched with a bimodal sin(i)~$\times$~Gaussian inclination distribution with width $\sigma_i = 6^{\circ}$ and $18^{\circ}$, respectively. The bimodal Rayleigh eccentricity distribution with width $\sigma_e = 0.02$ and $0.05$ provides a fine-match with the PS1 and ossos+ detections.

\begin{figure}
\includegraphics[width = \columnwidth]{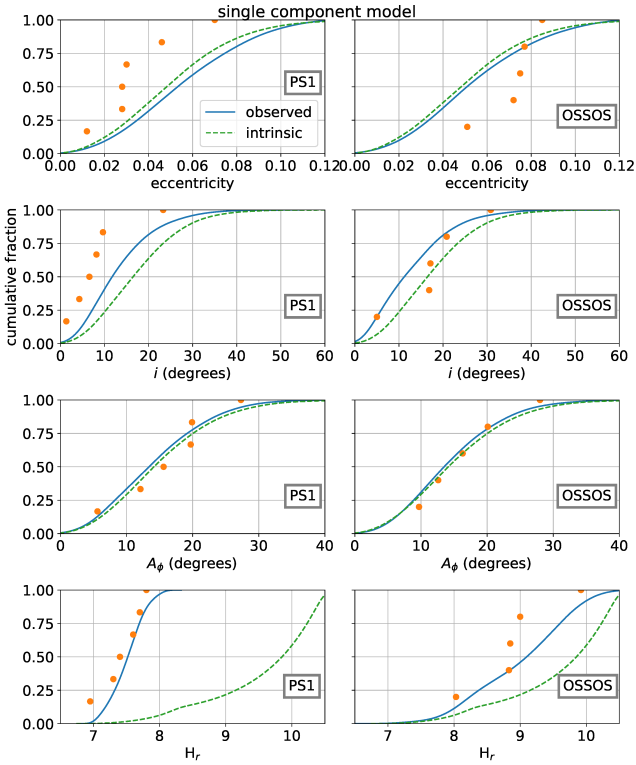}
\caption{Same as Figure~\ref{sim_vs_obs}, but modeling by a single component model.} 
\label{sim_vs_obs_single}
\end{figure}

Fig~\ref{sim_vs_obs_single} shows the cumulative distributions of synthetic detections generated from the single-component model. While the single component model is non-rejectable, it shows the difficultly to reproduce several observational results: the expected eccentricity distribution is either too high (PS1) or too low (OSSOS+) comparing with the observations, and while the inclination distribution seems well match with OSSOS+, it is too high for the PS1. Moreover, the single component model also expects to have more faint end detections (H$_r >$ 9) from OSSOS+, which is not the case. We summarize the parameters of our NT models in Table~\ref{tab:model}.

\begin{table*}
\caption{Neptunian Trojan Population Models \label{tab:model}}
\begin{tabular*}{\textwidth}{l @{\extracolsep{\fill}}lcccc}
\hline
& & \multicolumn{1}{c}{single-component model} &\multicolumn{2}{c}{two-component model} \\
\cline{4-5}
Parameter & Distribution  &  & cold & hot\\

\hline
\hline
$a$ (au) & constant & 30.1 & 30.1  & 30.1 &\\
$e$ & Rayleigh & $\sigma_e$ = 0.045  & $\sigma_e$ = 0.02 & $\sigma_e$ = 0.05\\
$i$  ($^{\circ}$) & $\sin$ i $\times$ Gaussian  &  $\sigma_i$ = 14$^a$   &  $\sigma_i$ = 6 &  $\sigma_i$ = 18 \\
$\omega$  ($^{\circ}$) & uniform & 0 - 2$\pi$  & 0 - 2$\pi$ & 0 - 2$\pi$\\ 
$\Omega$  ($^{\circ}$) & uniform & 0 - 2$\pi$ & 0 - 2$\pi$ & 0 - 2$\pi$\\ 
$M$  ($^{\circ}$)) & uniform & 0 - 2$\pi$  & 0 - 2$\pi$ & 0 - 2$\pi$\\ 
$\phi_{mean}$  ($^{\circ}$) & constant & 60  & 60 & 60\\
$A_{\phi_{1:1}}$  ($^{\circ}$) & Rayleigh & $\sigma_{A_{\phi}}$ = 12 &$\sigma_{A_{\phi}}$ = 12 & $\sigma_{A_{\phi}}$ = 12 \\
H$_r$  & power law & divot (0.9, 0.5, 3.2, 8.3)$^b$  &$\alpha$ = 0.2, cutoff at H$_r$=8 & divot (0.9, 0.5, 3.2, 8.3)$^b$ \\

\hline
\end{tabular*}
\\
$^a$  truncated at $i = 60^{\circ}$.\\
$^b$ ($\alpha_b$, $\alpha_f$, c, H$_b$), $\alpha_b$: bright end power law index, $\alpha_f$: faint end power law index, c: contrast between bright and faint end, H$_b$: break point. See \citet{law18b} for details.
\end{table*}

\subsection{The Population of Stable L4 Neptune Trojans}
\label{sec:pop}

With the bimodal inclination distribution from Sec.~\ref{sec:twodistrib}, we use the survey simulator to estimate the intrinsic number of both `cold' and `hot' stable L4 NTs. 
This is the total number of objects in the model population necessary to match the 6 and 5 detections of PS1 and OSSOS+, respectively. 
The absolute magnitude distribution of `cold' component is the same as Sec.~\ref{sec:twodistrib} with $\alpha = 0.2$ and a cutoff at $H_r = 8$.
For the absolute magnitude distribution of `hot' component, we use the parameters in \citet{law18b} and set $N(< H) \propto 10^{\alpha_b H_r}$ with a bright end slope $\alpha_b = 0.9$ from the largest observed NTs, H = 6.9, down to H = 8.3. We set a `divot' \citep{sha13, sha16} at H = 8.3, and after that we have a faint-end slope $\alpha_f = 0.5$ to satisfy the fact of no NT detected between $8.1 < $H$_r < 8.6$.


To match the number of $H_r < 8$ L4 NTs that were detected in our surveys (6 NTs, all from PS1), we require the L4 island to have a total $23^{+20}_{-11}$ NTs for $H_r < 8.0$, with $13^{+11}_{-6} $ and $10^{+9}_{-5}$ NTs from a cold and hot component, respectively.
To match the total number of L4 NTs that were detected in both of the surveys, which have $H_r < 10$ (roughly equal to a diameter of 50 km for an albedo of 0.05), the total L4 NT population should be $149^{+95}_{-81}$. Because the L4 NTs have no cold members $H_r \geq 8.0$, $136^{+84}_{-75}$ of the 149 NTs with $H_r < 10$ belong to hot component. This result is in good agreement with the result of \citet{lin19} of $162 \pm 73$, the more approximate estimate of 250 \citet{she10}, and the upper limit of 300 L4 NTs estimated by \citet{gla12}. 

On the other hand, if we consider the single-component Neptunian Trojan model, the total population estimation of L4 NT is $180^{+115}_{-98}$, which is close to the estimation using the two-component model.

Moreover, the MPC (Minor Planet Center) lists 63 L4 Jovian Trojans with $H_V < 10.2$ (2020 Feb. 7), which is equal to $H_r < 10$ if assuming the V-r color of Trojan is 0.2 \citep{jew18, smi02}. Thus, our result indicates that the L4 NT population is about 2.4 times greater than that of the L4 Jovian Trojans, for $H_r < 10$.

\subsection{Non-Detection of L5 NTs and Total Population}

The OSSOS+ survey did not detect any L5 NTs. However, Figure~\ref{fig:skyarea} shows that two OSSOS survey regions and the CFEPS survey overlaps with the L5 NT region. Could the OSSOS+ non-detection of L5 Trojans mean the L5 population is smaller than the L4 population?

Assuming that the populations at L4 and L5 are equal and symmetric, the expected number of detected L5 NTs is only $8\%$ of the total detections of simulated NTs. Thus, the non-detection of L5 NTs by OSSOS+ is expected: there is approximately a $67\%$ chance to have this non-detection result. 
We conclude that the assumption of same-sized populations in the L4 and L5 camps is not rejectable by the current NT sample. Such a result consists of the similar populations in the L4 and L5 clouds suggested by \citet{she10b}. Note that most of the simulated L5 NT detections belong to the cold component; without cold NT component in L5 island, the chances of a non-detection result in OSSOS+ are much higher.

If the L4 and L5 camps are symmetric, the total population estimation using two-component model will be: number of cold Trojans $= 26^{+22}_{-12} $ for $H_r < 8.0$, and number of hot Trojans is $272^{+168}_{-150}$ for $H_r < 10$. By using single-component model, we estimate the total number of NT population is $360^{+230}_{-196}$ for $H_r < 10$. Both results agree with the estimation of \citet{lin19}.

\section{Discussion}\label{sec:dis}

Since the two-component model is our preferred model for modeling the NT population, in the section, we discuss the possible scenarios for the formation of the two components.

In Section~\ref{sec:amp}, we found that the PS1 and OSSOS+ NTs have consistent libration amplitude distributions. Since the current libration amplitude distribution of NTs correlates to how they were captured \citep{fle00, lyk09, gom16}, such a result suggests that if the NT population has two components (see Section~\ref{sec:twodistrib}), they were likely captured at the same stage of Neptune's migration. 
Thus, we propose the following three possible scenarios:

{\bf 1. Formed as two components:} To form a `hot' component of NTs with $\sigma_i \sim 20 ^{\circ}$, the NTs need to be captured from an initially widely dispersed disk \citep{par15}. On the other hand, the `cold' component has only $\sigma_i \sim 6^\circ$ and can be captured from a thin disk \citep{nes09, lyk10, par15, che16}. 
If Neptune swept through a disk with both a thin and thick component, that might capture a population with two inclination distributions but one libration amplitude distribution. 
We know that the current classical Kuiper belt has two such overlapping populations \citep[though it is generally thought that the hot and cold components formed at separate locations in the original disk; see, e.g.][]{mor20}. 
Was such an overlapping population in place in the early, closer part of the disk?

A more likely two-component formation scenario is that, like \citet{lyk09, lyk11} discuss in detail, the origin and evolution of Trojans formed locally with Neptune (referred to in those papers as "pre-formed") and were captured from trans-Neptunian disks. So, there remains the possibility that a small fraction of local NTs survived to this date. Unsurprisingly, in general local NTs display low e, i ($<0.1$; $<5-10^\circ$), while their captured counterparts display wide ranges of e, i. Depending on the initial conditions or model details, the surviving fraction of local NTs may range from virtually zero to tens of $\%$ (\citet{che16} found similar results). Thus, it's possible that local and captured NTs survived with similar fractions and that they may be akin to cold/hot components in the cloud. This scenario has a observable consequence: we would expect the same cold/hot NT components can be also observed in L5 cloud. This scenario might also imply that the color distributions could differ between the cold and hot components, since the corresponding origins of the two components could have different color distributions.

{\bf 2. Formed as one component, evolved into two:}  This scenario originated in \citet{lin19}. The NT population could form with an intermediate-width inclination distribution and then evolve into two components.
\citet{alm09} found that the collision rate between Trojans and Plutinos is much higher than Plutino-Plutino or Trojan-Trojan collisions, and it is more effective for the low inclination objects. \citet{alm09} also suggested that this finding could explain the existence of size- and color-inclination dependencies in the Plutino population. If this is true, the same size- and color-inclination dependencies should also present in NT population. Especially since the Plutino population has a lower inclination distribution than the NT population, the high inclination NTs have higher chance to avoid collisions. 
This scenario may also explain why the colors of NTs differ from the color bimodality of the other trans-Neptunian object populations: the collisions remove the ultra-red matter \citep {gil02, gru09} from the surfaces of NTs. 
However, would the Trojan-Plutino collision rate be high enough? Can it eliminate the small-sized NTs to produce the cutoff after H$_r > 8$? This is also questionable, and beyond the scope of this paper.

{\bf 3. A collisional NT family in low eccentricity/inclination orbital space:} The only known collisional family in Kuiper Belt, the Haumea family, has a shallow H-distribution slope and lacks small family members \citep{pik20}. These facts suggest that the Haumea family formed in a graze-and-merge scenario rather than a catastrophic collision. Similar to the Haumea family, the cold component of NTs also has a shallow H-distribution slope and lacks members smaller than H$_r > 8$. Moreover, all of the cold component candidates (inclination $< 10^{\circ}$ and H$_r < 8$) have the same color \citep{jew18}. Could the cold component NTs belong to a collisional NT family in low eccentricity/inclination orbital space? Unlike scenario 1, there would likely be an asymmetry between the L4 and L5 NT populations, since there may not exist another collisional family in the L5 population. Such a consequence can be tested by future L5 NT surveys.

\section{Summary} 
\label{sec:sum}

We present the orbital properties of four newly discovered NTs by the near-ecliptic survey OSSOS. Three of them have orbital inclination $\gtrsim 17^{\circ}$, as expected for the dynamically hot inclination distribution of the NT population. Our numerical integrations for the four new OSSOS NTs showed that they are long-term dynamically stable in the 1:1 resonance, with two showing metastability  within their orbit fit uncertainty ranges.   

We explored the intrinsic libration amplitude, eccentricity, and inclination distributions of the stable NT population, using both the OSSOS+ surveys and the Pan-STARRS1 survey, via a survey simulator. 
Combined with an NT found earlier by \citet{ale16}, there are five NTs discovered by OSSOS+, and six stable NTs from the PS1 survey. 
The libration amplitude distribution can be described as a Rayleigh distribution with libration amplitude width $\sigma_{A_{\phi}} > 5\degr$. 
The best matched $\sigma_{A_{\phi}}$ is $15^\circ$ for both PS1 and OSSOS+. Using a Rayleigh eccentricity distribution model, the acceptable eccentricity width ($\sigma_e$) for both surveys is $\sim 0.045$. For a sin(i)~$\times$~Gaussian inclination distribution model, the acceptable inclination width ($\sigma_i$) for both surveys is 12--16\degr.

Considering the two surveys have very different magnitude limits and latitude coverage: OSSOS+ is much deeper than PS1 and focuses on the ecliptic.
The overlapping acceptable region for the eccentricity and inclination distributions derived from each survey are small and near the rejectable level, so we also consider an alternative scenario.

We propose size-dependent and bimodal eccentricity and inclination distributions for the stable NT population to explain the detections of NTs in the surveys we considered. One group, dynamically `cold' NTs, has a shallow H distribution with slope $\sim0.2$, and only contain large NTs ($H_r<8$) on low eccentricity and low inclination orbits with $\sigma_e \sim 0.02$ and $\sigma_i \sim 6^{\circ}$, respectively. The other group, dynamically `hot' NTs, have a wider range of sizes and eccentricity and inclination width of  $\sigma_e \sim 0.05$ and $\sigma_i \sim 18^{\circ}$. We perform a Bayesian model comparison to find the preferred model between this more complex two-component model and the simpler single-component model. The result shows substantial evidence that the two-component model is the preferred model to describe Neptunian Trojan population.

With the two-component NT population model, we found that there are $13^{+11}_{-6}$ `cold' NTs with $H_r < 8.0$, and $136^{+84}_{-75}$ `hot' NTs with $H_r < 10.0$ in the L4 island. On the other hand, the population of L4 NTs is $180^{+115}_{-98}$ if we use the single-component model. This result suggests that the NT population is 2.5 to 3 times larger than that of the Jovian Trojans within the same size range.

Although OSSOS has completed its observing, PS1 and now PS2 continue surveying and may discover more bright NTs. The Dark Energy Survey with CTIO has detected NTs \citep{ger16, lin19, bern19, kha20}, including those that are too small for PS1 to detect. 
Future faint NT detections will be tremendously enhanced by the discoveries of the Vera C. Rubin Observatory (LSST), particularly with the proposed North Ecliptic Spur survey \citep{ols18, sch18}.
These new bright- and faint-end NT samples will test the size-dependent bimodal NT eccentricity/inclination distribution that we propose.

\acknowledgments

The authors thank Matthew Holman, who kindly let us use the pointing records of the PS1 survey in this study. The authors also acknowledge the anonymous referees' useful suggestions for improving the manuscript.

HWL acknowledges the support of NASA Grant: NNX17AF21G.
KV acknowledges support from NASA (grants NNX15AH59G and 80NSSC19K0785) and NSF (grant AST-1824869).
M.T.B. appreciates support from UK STFC grant ST/P0003094/1 and travel support provided by STFC for UK participation in LSST through grant ST/N002512/1.
WHI acknowledges the support from MOST Grant: MOST 104-2119-008-024 (TANGO II), MOE under the Aim for Top University Program NCU, Macau Technical Fund: 017/2014/A1 and 039/2013/A2.

This research was supported by funding from the National Research Council of Canada and the National Science and Engineering Research Council of Canada. Based on observations obtained with MegaPrime/MegaCam, a joint project of the Canada-France-Hawaii Telescope (CFHT) and CEA/DAPNIA, at CFHT which is operated by the National Research Council (NRC) of Canada, the Institute National des Sciences de l’universe of the Centre National de la Recherche Scientifique (CNRS) of France, and the University of Hawaii, with this project receiving additional access due to contributions from the Institute of Astronomy and Astrophysics, Academia Sinica, National Tsing Hua University, and Ministry of Science and Technology, Taiwan.

The Pan-STARRS1 Surveys (PS1) have been made possible through contributions of the Institute for Astronomy, the University of Hawaii, the Pan-STARRS Project Office, the Max-Planck Society and its participating institutes, the Max Planck Institute for Astronomy, Heidelberg and the Max Planck Institute for Extraterrestrial Physics, Garching, The Johns Hopkins University, Durham University, the University of Edinburgh, Queen's University Belfast, the Harvard-Smithsonian Center for Astrophysics, the Las Cumbres Observatory Global Telescope Network Incorporated, the National Central University of Taiwan, the Space Telescope Science Institute, the National Aeronautics and Space Administration under Grant No. NNX08AR22G issued through the Planetary Science Division of the NASA Science Mission Directorate, the National Science Foundation under Grant No. AST-1238877, and the University of Maryland.

%

\vspace{5mm}
\facilities{CFHT (MegaPrime), PS1}
\software{orbfit \citep{ber00}, SWIFT \citep{lev94}, OSSOS Survey Simulator \citep{ban16, law18a}}



\end{document}